\documentclass[a4paper]{jpconf}
\usepackage{graphicx}

\begin{document}
\title{Barrier modification in sub-barrier fusion reactions using
Wong formula with Skyrme forces in semiclassical formalism}

\author{Raj Kumar and Raj K. Gupta}

\address{Physics Department, Panjab University, Chandigarh-160014, INDIA.}


\begin{abstract}
We obtain the nuclear proximity potential by using semiclassical
extended Thomas Fermi (ETF) approach in Skyrme energy density
formalism (SEDF), and use it in the extended $\ell$-summed Wong
formula under frozen density approximation. This method has the
advantage of allowing the use of different Skyrme forces, giving
different barriers. Thus, for a given reaction, we could choose a
Skyrme force with proper barrier characteristics, not-requiring
extra ``barrier lowering" or ``barrier narrowing" for a best fit
to data. For the $^{64}$Ni+$^{100}$Mo reaction, the $\ell$-summed
Wong formula, with effects of deformations and orientations of
nuclei included, fits the fusion-evaporation cross section data
exactly for the force GSkI, requiring additional barrier
modifications for forces SIII and SV. However, the same for other
similar reactions, like $^{58,64}$Ni+$^{58,64}$Ni, fits the data
best for SIII force. Hence, the barrier modification effects in
$\ell$-summed Wong expression depends on the choice of Skyrme
force in extended ETF method.
\end{abstract}

\section{Introduction}
The unexpected behavior of some fusion-evaporation cross sections
at energies far below the Coulomb barrier, has challenged the
theoretical models to explain the, so called, fusion hinderance
phenomena in true coupled-channels calculations (ccc) for
reactions such as $^{58}$Ni+$^{58}$Ni, $^{64}$Ni+$^{64}$Ni, and
$^{64}$Ni+$^{100}$Mo \cite{jiang04}. The ccc could, however, be
sensitive to the so far unobserved, hence not-included, high-lying
states. Misicu and Esbensen \cite{misicu06} were the first who
succeeded in describing the above said three reactions in terms of
a density-dependent M3Y interaction, modified by adding a
repulsive core potential \cite{misicu04}. The repulsive core
changes the shape of the inner part of the potential in terms of a
thicker barrier (reduced curvature $\hbar\omega$) and shallower
pocket. Here, deformations are included up to hexadecapole
($\beta_2-\beta_4$) and the orientation degrees of freedom is
integrated over all the allowed values in the same plane.

The dynamical cluster-decay model (DCM) of preformed clusters by
Gupta and collaborators \cite{gupta08,gupta10} is found recently
\cite{arun09,gupta09} to have barrier modification effects as the
inbuilt property, where ``barrier lowering" at sub-barrier
energies arise in a natural way in its fitting of the only
parameter of model, the neck-length parameter. The difference of
actually-calculated barrier from the actually-used barrier height,
corresponding to the neck-length parameter for best-fitted
fusion-evaporation cross section, gives the ``barrier lowering" in
DCM, whose values are found to increase as the incident energy
decreases to sub-barrier energies. Calculations are based on
$\beta_2$ deformations and orientation $\theta_i$-dependent
nuclear proximity potential of Blocki {\it et al.}
\cite{blocki77}.

Very recently, the Wong formula \cite{wong73} is also extended by
Gupta and collaborators \cite{kumar09,bansal10} to include its
angular momentum $\ell$-summation explicitly, which is also shown
to contain the barrier modification effects due to the
$\ell$-dependent barriers. However, for the $^{58,64}$Ni-based
fusion-evaporation cross sections \cite{jiang04}, a further
modification of barriers is found essential for below-barrier
energies, which is implemented empirically either by ``lowering
the barrier" or ``narrowing the barrier curvature" by a fixed
amount for all $\ell$'s in the potential calculated by using the
proximity potential of Blocki {\it et al.} \cite{blocki77}, but
with multipole deformations $\beta_2-\beta_4$ and
$\theta_i$-integrated for co-planer nuclei. Apparently, the depth
of the potential pocket plays no role, in both the DCM and
$\ell$-summed Wong formula (the two models are same for capture
reactions).

In this contribution, we use within the $\ell$-summed Wong model,
the nuclear proximity potential obtained recently \cite{gupta09a}
for the Skyrme nucleus-nucleus interaction in the semiclassical
ETF approach. Using SEDF, the universal function of proximity
potential is obtained as a sum of the parametrized
spin-orbit-density-independent and the
spin-orbit-density-dependent universal functions (UF's), with
different parameters of UF's obtained for different Skyrme forces
\cite{gupta09a}. This method has the advantage of introducing the
barrier modifications at sub-barrier energies, if needed, by
either (i) modifying the Fermi density parameters (the
half-density radii and/ or surface thicknesses, for ``exact" SEDF
calculations \cite{gupta09a}), (ii) the constants of the
parametrized UF's \cite{gupta09a} or (iii) change the Skyrme force
itself since a different Skyrme force would give different barrier
characteristics (height and curvature). This later possibility is
exploited here in this paper. It is possible that some Skyrme
force would fit the data for one reaction, but not for another
reaction and hence requiring ``barrier modification" or another
Skyrme force.

Section 2 gives briefly the semiclassical ETF method using SEDF,
including details of approximations used for adding densities.
Section 3 discusses the $\ell$-summed Wong formula \cite{kumar09}.
Our calculations are given in section 4, and a brief summary of
results in section 5.

\section{\label{sec:level2}The semiclassical extended Thomas Fermi (ETF) model}
The SEDF in semiclassical ETF method provides a convenient way for
calculating the interaction potential between two nuclei. In the
Hamiltnian density, the kinetic energy density $\tau$ as well as
the spin-orbit density $\vec{J}$ are functions of the nucleon
density $\rho_q$, $q=n,p$. For the composite system, the densities
can be added in either adiabatic or sudden approximation, but we
are interested in sudden densites since the different terms of
Skyrme Hamiltonian density are then found to constitute the
nuclear proximity potential \cite{gupta09a,gupta07,chatto84}. The
sudden densities are defined with or without exchange effects (due
to anti-symmetrization), and the one without exchange effects is
also refered to as frozen density \cite{li91}. In ETF method, the
lowest order $\tau$ is the Thomas Fermi (TF) kinetic energy
density $\tau_{TF}$, which already contains a large part of the
exchange effects, and that the higher order terms include exchange
effects in full. Here we limit $\tau(\vec{r})$ and
$\vec{J}(\vec{r})$ to second order terms  for reasons of being
enough for numerical convergence \cite{bartel02}.

The nucleus-nucleus interaction potential in SEDF, based on
semiclassical ETF model, is
\begin{small}
\begin{equation}
V_N(R)=E(R)-E({\infty})
=\int{H(\vec{r})d\vec{r}}-\left[\int{H_1(\vec{r})d\vec{r}}+\int{H_2(\vec{r})}d\vec{r}\right],
\label{eq:1}
\end{equation}
\end{small}
where the Skyrme Hamiltonian density
\begin{small}
\begin{eqnarray}
H(\rho,\tau,\vec{J})&=&\frac{\hbar^{2}}{2m}\tau
+\frac{1}{2}t_{0}\left[(1+\frac{1}{2}x_{0})\rho^2-(x_{0}+\frac{1}{2})
(\rho_n^2+\rho_p^2)\right] \nonumber\\
&&+\frac{1}{12}t_3\rho^\alpha\left[(1+\frac{1}{2}x_3)\rho^2-(x_3+\frac{1}{2})
(\rho_n^2+\rho_p^2)\right]
+\frac{1}{4}\left[t_1(1+\frac{1}{2}x_1)+t_2(1+\frac{1}{2}x_2)\right]
\rho\tau \nonumber\\
&&-\frac{1}{4}\left[t_1(x_1+\frac{1}{2})-t_2(x_2+\frac{1}{2})\right]
(\rho_n\tau_n+\rho_p\tau_p)
+\frac{1}{16}\left[3t_1(1+\frac{1}{2}x_1)-t_2(1+\frac{1}{2}x_2)\right]
(\vec{\nabla}\rho)^2 \nonumber\\
&&-\frac{1}{16}\left[3t_1(x_1+\frac{1}{2})+t_2(x_2+\frac{1}{2})\right]
\left[(\vec{\nabla}\rho_n)^2+(\vec{\nabla}\rho_p)^2\right] \nonumber\\
&&-\frac{1}{2}W_0\left[\rho\vec{\nabla}\cdot\vec{J}+\rho_n\vec{\nabla}\cdot\vec{J_n}
+\rho_p\vec{\nabla}\cdot\vec{J_p}\right] .
\label{eq:2}
\end{eqnarray}
\end{small}
\par\noindent
Here, $\rho=\rho_n+\rho_p$, $\tau=\tau_n+\tau_p$,
$\vec{J}=\vec{J}_n+\vec{J}_p$ are the nuclear, kinetic energy and
spin-orbit densities, respectively. m is the nucleon mass. $x_i$,
$t_i$ ($i$=0,1,2,3), $\alpha$ and $W_0$ are the Skyrme force
parameters, fitted by different authors to ground state properties
of various nuclei (see, e.g., \cite{brack85,friedrich86}). Of the
available forces, we use the old, well known SIII and SV forces.
Coulomb effects are added directly. Recently, Agrawal {\it et.
al.} \cite{agrawal06} modified the Hamiltonian density
(\ref{eq:2}) on two accounts, and obtained a new force GSKI: (i)
the third term in (\ref{eq:2}) is replaced as
\begin{small}
\begin{equation}
\frac{1}{2}\sum_{i=1}^{3} t_{3i}\rho^{\alpha_i}\left[(1+\frac{1}{2}x_{3i})\rho^2-(x_{3i}+\frac{1}{2})
(\rho_n^2+\rho_p^2)\right] ,
\label{eq:3}
\end{equation}
\end{small}
and (ii) a new term due to tensor coupling with spin and gradient is added as
\begin{small}
\begin{equation}
-\frac{1}{16}(t_1x_1+t_2x_2)\vec{J}^2+\frac{1}{16}(t_1-t_2)(\vec{J}_p^2+\vec{J}_n^2).
\label{eq:4}
\end{equation}
\end{small}
We have also used this new GSKI force, with additional six, two
each of $x_{3}$, $t_{3}$ and $\alpha$, constants.

The kinetic energy density in ETF method, up to second order
\cite{bartel02}, for $q$=$n$ or $p$ is
\begin{small}
\begin{eqnarray}
\tau_q(\vec{r})&=&\frac{3}{5}(3\pi^2)^{2/3}\rho_q^{5/3}+\frac{1}{36}\frac{(\vec{\nabla}\rho_q)^2}{\rho_q}
+\frac{1}{3}\Delta\rho_q+\frac{1}{6}\frac{\vec{\nabla}\rho_q\cdot \vec{\nabla}f_q+\rho_q\Delta f_q}{f_q} \nonumber\\
&&-\frac{1}{12}\rho_q \left(\frac{\vec{\nabla}f_q}{f_q}\right)^2+
\frac{1}{2}\rho_q\left(\frac{2m}{\hbar^2}\right)^2\left(\frac{W_0}{2}
\frac{\vec{\nabla}(\rho+\rho_q)}{f_q}\right)^2,
\label{eq:5}
\end{eqnarray}
\end{small}
with $f_q$ as the effective mass form factor,
\begin{small}
\begin{equation}
f_q(\vec{r})=1+\frac{2m}{\hbar^2}\frac{1}{4}\left\{ t_1(1+\frac{x_1}{2})+t_2(1+\frac{x_2}{2}) \right\} \rho(\vec{r})
-\frac{2m}{\hbar^2}\frac{1}{4}\left\{ t_1(x_1+\frac{1}{2})-t_2(x_2+\frac{1}{2}) \right\}\rho_q(\vec{r}).
\label{eq:6}
\end{equation}
\end{small}
Note that both $\tau_q$ and $f_q$ are each functions of $\rho_q$ and/ or $\rho$ only.

The spin $\vec{J}$ is a purely quantal property, and hence has no
contribution in the lowest (TF) order. However, at the ETF level,
the second order contribution gives
\begin{small}
\begin{equation}
\vec{J_q}(\vec{r})=-\frac{2m}{\hbar^2}\frac{1}{2}W_0\frac{1}{f_q}
\rho_q\vec{\nabla}(\rho+\rho_q).
\label{eq:7}
\end{equation}
\end{small}
\par\noindent
Note, $\vec{J_q}$ is also a function of $\rho_q$ and/ or $\rho$ alone.

Next, for the proximity potential we introduce the slab
approximation of semi-infinite nuclear matter with surfaces
parallel to $x-y$ plane, moving in $z$-direction, and separated by
distance $s$ having minimum value $s_0$. Then, following Blocki
{\it et al.} \cite{blocki77} and Gupta {\it et al.}
\cite{gupta09a,gupta07,chatto84}, the interaction potential
$V_N(R)$ between two nuclei separated by $R=R_{1}+R_{2}+s$, is
given as
\begin{eqnarray}
V_N(R)&=&2\pi\bar{R}\int_{s_0}^{\infty}{e(s)ds} \nonumber\\
&=&2\pi\bar{R}\int
\left\{H(\rho,\tau,\vec{J})-[H_1(\rho_1,\tau_1,\vec{J_1})+H_2(\rho_2,\tau_2,\vec{J_2})]\right\}dz \nonumber\\
&=&4\pi\bar{R}\gamma b \phi(D).
\label{eq:8}
\end{eqnarray}
where $\bar{R}$ is the mean curvature radius, and $e(s)$ is the
interaction energy per unit area between the flat slabs giving the
universal function $\phi(D)$ in terms of a dimensionless variable
$D=s/b$, with sufrace width $b$=0.99 fm. The nuclear surface
energy constant $\gamma=0.9517\big [1-1.7826\big
(\frac{N-Z}{A}\big )^2\big ]$ MeV fm$^{-2}$. $\phi(D)$ can be
calculated ``exactly" or parametrized in terms of exponential
and/or polynomial functions \cite{gupta09a}. For axially deformed
and oriented nuclei, $\bar{R}$ is given in terms of the radii of
curvature $R_{i1}$ and $R_{i2}$ in the principal planes of
curvature of each of the two nuclei ($i$=1,2) at the points of
closest approach (defining $s_0$), by
\begin{equation}
{{1}\over {\bar{R}^2}}={{1}\over {R_{11}R_{12}}} +{{1}\over {R_{21}R_{22}}}
+\left [{{{1}\over {R_{11}R_{21}}}+{{1}\over {R_{12}R_{22}}}} \right ]sin^2\Phi
+\left [{{{1}\over {R_{11}R_{22}}}+{{1}\over {R_{21}R_{12}}}}\right ]cos^2\Phi.
\label{eq: 9}
\end{equation}
Here, $\Phi$ is the azimuthal angle between the principal planes
of curvature of two nuclei ($\Phi$=$0^0$ for co-planar nuclei).
The four principal radii of curvature are given in terms of radii
$R_i({\alpha_i})$ and their first and second order derivatives
$R_i^{\prime}({\alpha_i})$ and $R_i^{\prime\prime}(\alpha_i)$ {\it
w.r.t.} ${\alpha_i}$, where
\begin{equation}
R_i(\alpha_i)=R_{0i}\big [1+\sum_{\lambda}\beta _{{\lambda}i}Y_{\lambda}^{(0)} (\alpha_i)\big ],
\label{eq:10}
\end{equation}
with $R_{0i}$ as the spherical or half-density nuclear radius,
$\lambda$=2,3,4..., as multipole deformations, and $\alpha _i$ as
an angle between radius vector $R_i({\alpha_i})$ and symmetry
axis, measured clockwise from symmetry axis. For the estimation of
$s_0$, we refer to \cite{gupta04} for $\Phi$=0 and to
\cite{manhas05} for $\Phi\ne$0.

For nuclear denesity $\rho_i$ of each nucleus ($i$=1,2), we use
the temperature T-dependent, two-parameter Fermi density (FD)
distribution, which for the slab approximation is given by
\begin{equation}
\rho_i(z_i)=\rho_{0i}(T)\left[1+\exp\left(\frac{z_i-R_{i}(T)}{a_{i}(T)}\right)\right]^{-1} -\infty\leq z\leq \infty
\label{eq:11}
\end{equation}
with $z_2=R-z_1=(R_1(\alpha_1)+R_2(\alpha_2)+s)-z_1$, and central density
$\rho_{0i}(T)=\frac{3A_i}{4 \pi R_{i}^3(T)}\big [1+\frac{\pi^2a_{i}^2(T)}{R_{i}^2(T)}\big ]^{-1}$.
Then, since $\rho_i=\rho_{n_i}+\rho_{p_i}$, following our earlier work \cite{gupta09a}, for nuclen density we define
\begin{equation}
\rho_{n_i}=(N_i/A_i)\rho_i \qquad \hbox{and} \qquad \rho_{p_i}=(Z_i/A_i)\rho_i,
\label{eq:12}
\end{equation}
with half density radii $R_{0i}$ and surface thickness parameters $a_{0i}$ in Eq. (\ref{eq:11}) at $T$=0, obtained by
fitting the experimental data to respective polynomials in nuclear mass region $A$=4-238, as
\begin{equation}
R_{0i}(T=0)=0.9543+0.0994A_i-9.8851\times10^{-4}A_i^2+4.8399\times10^{-6}A_i^3-8.4366\times10^{-9}A_i^4
\label{eq:13}
\end{equation}
\begin{equation}
a_{0i}(T=0)=0.3719+0.0086A_i-1.1898\times10^{-4}A_i^2+6.1678\times10^{-7}A_i^3-1.0721\times10^{-9}A_i^4.
\label{eq:14}
\end{equation}
The T-dependence in the above formulas are then introduced as in Ref. \cite{shlomo91},
\begin{equation}
R_{0i}(T)=R_{0i}(T=0)[1+0.0005T^2],\qquad a_{0i}(T)=a_{0i}(T=0)[1+0.01T^2].
\label{eq:15}
\end{equation}
Also, the surface width $b$ is made T-dependent \cite{royer92}, $b(T)=0.99(1+0.009T^2)$, where T is related to the incoming
center-of-mass energy $E_{c.m.}$ or the compound nucleus excitation energy $E_{CN}^*$ via the entrance channel
$Q_{in}$-value, as $E_{CN}^{\ast }=E_{c.m.}+Q_{in}={1\over 9}AT^2-T \quad \hbox{($T$ in MeV)}$.

Next, for the composite system, $\rho=\rho_1+\rho_2$, and the $\tau(\rho)$ and $\vec{J}(\rho)$ are added as per prescrption
used. For sudden approximation (with exchange effects),
\begin{small}
\begin{equation}
\tau(\rho)=\tau(\rho_{1n}+\rho_{2n})+\tau(\rho_{1p}+\rho_{2p}), \qquad \hbox{and} \qquad
\vec{J}(\rho)=\vec{J}(\rho_{1n}+\rho_{2n})+\vec{J}(\rho_{1p}+\rho_{2p}),
\label{eq:16}
\end{equation}
\end{small}
and for the frozen approximation (equivalently, sudden without exchange effects),
\begin{equation}
\tau(\rho)=\tau_1(\rho_1)+\tau_2(\rho_2),\qquad \hbox{and} \qquad
\vec{J}(\rho)=\vec{J}_1(\rho_1)+\vec{J}_2(\rho_2),
\label{eq:17}
\end{equation}
with $\rho_i=\rho_{in}+\rho_{ip}$,
$\tau_i(\rho_i)=\tau_{in}(\rho_{in})+\tau_{ip}(\rho_{ip})$, and
$\vec{J}_i(\rho_i)=\vec{J}_{in}(\rho_{in})+\vec{J}_{ip}(\rho_{ip})$.
In the following, we consider only frozen densities (sudden
without exchange), since $\ell$-summed Wong formula within sudden
approximation does not fit the $^{64}$Ni-based reactions data for
use of SIII force \cite{kumar09a}.

Finally, adding the Coulomb and centrifugal interactions to the
nuclear interaction potential $V_N(R)$, we get the total
interaction potential for deformed and oriented nuclei
\cite{gupta08,gupta05}, as
\begin{equation}
V_{\ell}(R)=V_N(R,A_{i},\beta_{\lambda i},T,\theta_{i},\Phi)+V_{C}(R,Z_{i},\beta_{\lambda i},T,\theta_{i},\Phi)
+V_{\ell}(R,Z_{i},\beta_{\lambda i},T,\theta_{i},\Phi),
\label {eq:18}
\end{equation}
with non-sticking moment-of-inertia $I_{NS}$ (=$\mu R^2$) for
$V_{\ell}$. Eq.(\ref{eq:18}) gives the barrier height
$V_B^{\ell}$, position $R_B^{\ell}$, and the curvature
$\hbar\omega_{\ell}$ for each $\ell$, to be used in extended
$\ell$-summed Wong's formula \cite{kumar09}, discussed in the
following section.

\section{\label{sec:level3}Extended $\ell$-summed Wong Formula}
According to Wong \cite{wong73}, in terms of $\ell$ partial waves,
the fusion cross-section for two deformed and oriented nuclei
colliding with $E_{c.m.}$ is
\begin{equation}
\sigma (E_{c.m.},\theta_i,\Phi)=\frac{\pi}{k^2}\sum_{\ell=0}^{\ell_{max}}
(2\ell+1)P_{\ell}(E_{c.m.},\theta_i,\Phi), \qquad k=\sqrt{\frac{2\mu E_{c.m.}}{\hbar^2}}
\label{eq:19}
\end{equation}
with $\mu$ as the reduced mass. Here, $P_{\ell}$ is the
transmission coefficient for each $\ell$ which describes the
penetration of barrier $V_{\ell}(R,E_{c.m.},\theta_i,\Phi)$. Using
Hill-Wheeler \cite{hill53} approximation, the penetrability
$P_{\ell}$, in terms of its barrier height
$V_B^{\ell}(E_{c.m.},\theta_i,\Phi)$ and curvature
$\hbar\omega_{\ell}(E_{c.m.},\theta_i,\Phi)$, is
\begin{equation}
P_{\ell}=\left[1+\exp\left(\frac{2\pi(V_B^{\ell}(E_{c.m.},\theta_i,\Phi)-E_{c.m.})}
{\hbar\omega_{\ell}(E_{c.m.},\theta_i,\Phi)}\right)\right]^{-1},
\label{eq:20}
\end{equation}
with $\hbar\omega_{\ell}$ evaluated at the barrier position
$R=R_B^{\ell}$ corresponding to $V_B^{\ell}$. Note, the
$\ell$-dependent potentials are required here, given by
Eq.(\ref{eq:18}). Carrying out the $\ell$-summation in
Eq.(\ref{eq:19}) empirically for a best fit to measured
cross-section \cite{kumar09}, and on integrating over the angles
$\theta_i$ and $\Phi$, we get the fusion cross-section
$\sigma(E_{c.m.})$.

\section{Calculations}
We have made our calculations for the $^{64}$Ni+$^{100}$Mo
reaction, using SIII, SV and GSKI forces, with frozen densities.
Fig. 1(a) shows for one $E_{c.m.}$ and fixed ($\theta_i$, $\Phi$),
a comparison of interaction potentials for the three forces,
illustrating their barrier height and position to be
force-dependent. This is an interesting property, which we use to
fit the fusion-evaporation cross-section in Fig. 1(b). Interesting
enough, the data \cite{jiang04} for the above said
$^{64}$Ni+$^{100}$Mo reaction fit the $\theta_i$-integrated
($\Phi$=0$^0$) $\ell$-summed Wong formula for only the new force
GSKI. The other forces (SIII and SV) would apparently need
additional barrier modification effects to be added empirically
\cite{kumar09}. From the deduced $\ell_{max}$-values, presented in
Fig. 1(c), we notice that $\ell_{max}$ as a function of $E_{c.m.}$
vary smoothly only for GSKI force, achieving zero value at
sub-barrier energies and a tendency to saturate at an
above-barrier energy. For the other forces (SIII and SV),
non-fitting the data, the $\ell_{max}$ varies with $E_{c.m.}$
erratically, in particular at sub-barrier energies, but could also
be smoothed by adding futher ``barrier lowering" or ``barrier
narrowing" emprically \cite{kumar09}.

\section{Summary and discussion}
Concluding, the $\ell$-summed Wong expression, using the barriers
calculated in frozen-density approximation in semiclassical
extended Thomas Fermi method, based on Skyrme energy density
formalism, describes the fusion-evaporation cross-section data for
$^{64}$Ni+$^{100}$Mo reaction nicely with the new Syrme force GSKI
only. The variation of deduced $\ell_{max}$ with $E_{c.m.}$ is
found smooth. Other Skyrme forces (SIII and SV) demand additional
barrier modifications at sub-barrier energies for this reaction.
However, the same calculation when applied \cite{kumar09a} to
$^{58,64}$Ni+$^{58,64}$Ni data result in a similar good fit, with
a smooth dependence of $\ell_{max}$ on $E_{c.m.}$, for only the
Skyrme force SIII. Thus, barrier-modification or no
barrier-modification in $\ell$-summed Wong expression depends on
the choice of Skyrme force in semiclassical ETF method for frozen
densities.

\begin{figure}[h]
\begin{center}
\vspace{-1pc}
\includegraphics[width=12pc]{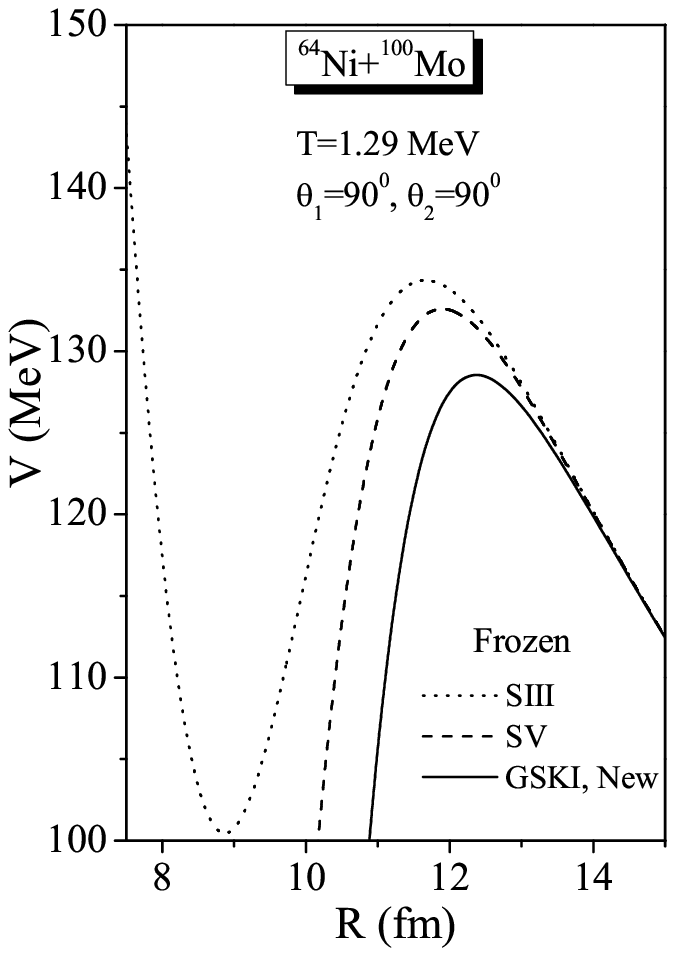}
\includegraphics[width=12pc]{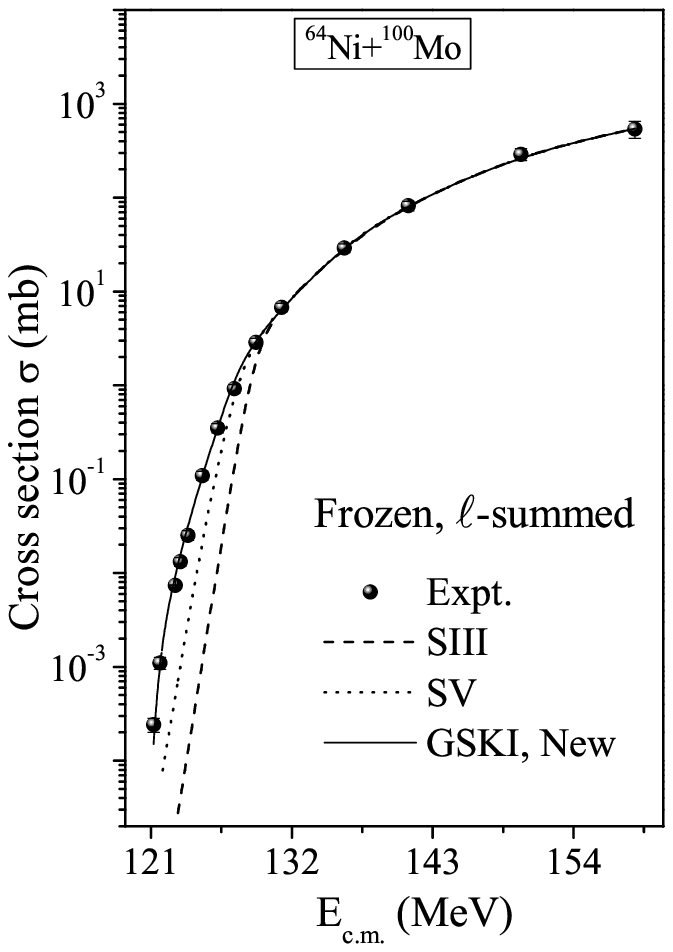}
\includegraphics[width=12pc]{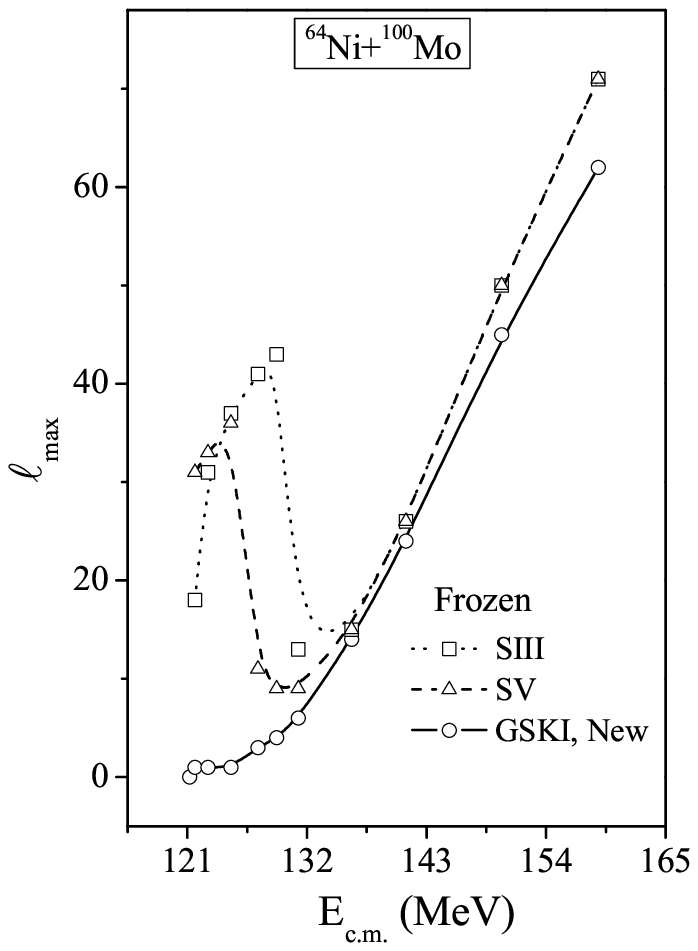}
\vspace{-0.5pc} \caption{\label{fig:1}(a) Interaction potentials
for one $E_{c.m.}$ and fixed ($\theta_i$, $\Phi$), (b)
fusion-evaporation cross-section as a function of $E_{c.m.}$ for
$^{64}$Ni+$^{100}$Mo reaction, calculated by using $\ell$-summed
Wong formula integrated over $\theta_i$ ($\Phi$=0$^0$) and
compared with experimental data \cite{jiang04}, and (c) deduced
$\ell_{max}$ values {\it vs.} $E_{c.m.}$, for the Skyrme forces
SIII, SV and GSKI, using frozen densities. }
\end{center}
\end{figure}

\section{Acknowledgments}
The financial support from the Department of Sc. \& Tech. (DST),
Govt. of India, and Council of Sc. \& Industial Research (CSIR),
New Delhi is gratefully acknowledged.


\section*{References}
\begin{thebibliography}{99}
\bibitem{jiang04}
Jiang C L {\it et al.} 2004 \emph{Phys. Rev. Lett.} {\bf 93} 012701;
Jiang C L {\it et al.} 2005 \emph{Phys. Rev. C.} {\bf 71} 044613
\bibitem{misicu06}
Misicu S and Esbensen H 2006 \emph{Phys. Rev. Lett.} {\bf 96} 112701
\bibitem{misicu04}
Misicu S and Greiner W  2004 \emph{Phys. Rev. C} {\bf 69} 054601
\bibitem{gupta08}
See, e.g., the review: Gupta R K , Arun S K, Kumar R, and Niyti 2008 \emph{Int. Rev. Phys. (IREPHY)} {\bf 2} 369
\bibitem{gupta10}
Gupta R K 2010 \emph{Lecture Notes in Physics, ``Clusters in Nuclei"}, ed Beck C, Vol. I, (Springer Verlag) p 223
\bibitem{arun09}
Arun S K, Kumar R, and Gupta R K 2009 \emph{J. Phys. G: Nucl. Part. Phys.} {\bf 36} 085105
\bibitem{gupta09}
Gupta R K, Arun S K, Kumar R and Bansal M 2009 \emph{Nucl. Phys. A} {\bf 834} 176c
\bibitem {blocki77}
Blocki J, Randrup J, Swiatecki W J, and Tsang C F 1977 \emph{Ann. Phys. (N.Y.)} {\bf 105} 427
\bibitem{wong73}
Wong C Y 1973 \emph{Phys. Rev. Lett.} {\bf 31}, 766
\bibitem{kumar09}
Kumar R, Bansal M, Arun S K, and Gupta R K 2009 \emph{Phys. Rev. C} {\bf 80} 034618
\bibitem{bansal10}
Bansal M and Gupta R K 2010 \emph{Phys. Rev. C}, to be published.
\bibitem{gupta09a}
Gupta R K, Singh D, Kumar R and Greiner W 2009 \emph{J. Phys. G: Nucl. Part. Phys.} {\bf 36} 075104
\bibitem{gupta07}
Gupta R K, Singh D, and Greiner W 2007 \emph{Phys. Rev. C} {\bf 75} 024603
\bibitem{chatto84}
Chattopadhyay P and Gupta R K 1984 \emph{Phys. Rev. C} {\bf30} 1191
\bibitem{li91}
Li G -Q 1991 \emph{J. Phys. G: Nucl. Part. Phys.} {\bf 17} 1
\bibitem{bartel02}
Bartel J and Bencheikh K 2002 \emph{Eur. Phys. J. A} {\bf 14} 179
\bibitem{brack85}
Brack M, Guet C, and Hakansson H -B 1985 \emph{Phys. Rep.} {\bf 123} 275
\bibitem{friedrich86}
Friedrich J and Reinhardt P -G 1986 \emph{Phys. Rev. C} {\bf 33} 335
\bibitem{agrawal06}
Agrawal B K, Dhiman S K, and Kumar R 2006 \emph{Phys. Rev. C.} {\bf 73} 034319
\bibitem{gupta04}
Gupta R K, Singh N, and Manhas M 2004 \emph{Phys. Rev. C} {\bf 70} 034608
\bibitem{manhas05}
Manhas M and Gupta R K 2005 \emph{Phys. Rev. C} {\bf 72} 024606
\bibitem{shlomo91}
Shlomo S and Natowitz J B 1991 \emph{Phys. Rev. C} {\bf 44} 2878
\bibitem{royer92}
Royer G and Mignen J 1992 \emph{J. Phys. G: Nucl. Part. Phys.} {\bf 18} 1781
\bibitem{kumar09a}
Kumar R and Gupta R K, {\it Proc. Int. Symp. on Nucl. Phys.} 2009 {\bf 54}, (BRNS-DAE, Govt. of India) p 288
\bibitem{gupta05}
Gupta R K {\it et al.}
2005 \emph{J. Phys. G: Nucl. Part. Phys.} {\bf 31} 631
\bibitem{hill53}
Hill D L and Wheeler J A 1959 \emph{Phys. Rev.} {\bf 89} 1102; Thomas T D 1959 \emph{Phys. Rev.} {\bf 116} 703.

\end{thebibliography}
\end{document}